\documentclass[reprint,aps,pra,twocolumn,superscriptaddress,
longbibliography,floatfix,secnumarabic,footinbib]{revtex4-1}%
\usepackage{graphicx}
\usepackage{amsfonts}
\usepackage{amsmath}
\usepackage{ltxtable}
\usepackage{CJK}
\usepackage{mdwmath}
\usepackage{color}
\usepackage{appendix}
\usepackage{mathtools}
\usepackage{amssymb}
\usepackage{bm}
\usepackage{bbold}
\usepackage{booktabs}
\usepackage{threeparttable}
\usepackage[normalem]{ulem}
\usepackage[colorlinks,allcolors=blue]{hyperref}

\newcommand{\mC}{{\mathcal C}}

\newcommand{\mS}{{\mathcal S}}
\newcommand{\mT}{{\mathcal T}}
\newcommand{\mI}{{\mathcal I}}
\newcommand{\mJ}{{\mathcal J}}

\begin{document}

\title{Hardy-like Quantum Pigeonhole Paradox and the Projected-Coloring Graph State}

\author{Weidong Tang}
\email{wdtang@snnu.edu.cn}
\affiliation{School of Mathematics and Statistics, Shaanxi Normal University, Xi'an 710119, China}

\begin{abstract}

A Hardy-like version of the quantum pigeonhole paradox is proposed, which can also be considered as a special kind of Hardy's paradox.  Besides an example induced from the minimal system, a general construction of this paradox from an $n$-qubit quantum state is also discussed. Moreover, by introducing the projected-coloring graph and the projected-coloring graph state, a pictorial representation of the Hardy-like quantum pigeonhole paradox can be presented. This Hardy-like version of quantum pigeonhole paradox can be implemented more directly in the experiment than the original one, since it does not require some sophisticated techniques such as weak measurements. In addition, from the angle of Hardy's paradox,  some Hardy-like quantum pigeonhole paradoxes can even set a new record for the success probability of demonstrating Bell nonlocality.

\end{abstract}

\maketitle

\section{Introduction}
When three two-level quantum particles(qubits) are pre- and postselected in a specific subensemble,  an effect that no two particles being in the same quantum state could arise, conflicting with the pigeonhole principle which states that if three pigeons are put into two boxes,  necessarily two pigeons will stay in the same box. Such counterintuitive quantum feature provides an interesting demonstration that quantum correlations cannot be simulated classically, and was referred to as quantum pigeonhole effect(or paradox) in the seminal work\cite{Aharonov2016} by Aharonov et al. In some sense, it can also be regraded as a demonstration of Bell nonlocality\cite{Bell} or contextuality\cite{KS,SixiaYu2014} without inequalities.

However, since quantum pigeonhole paradox is a kind of pre- and post-selection\cite{PPS1964,PPS2005} effect, usually it cannot be implemented directly and the experimental demonstration requires some sophisticated techniques such as weak-measurements\cite{Aharonov-AAV-1988,Waegell2017,MingChengChen2019}. To overcome this limitation, a natural thought is to explore new versions of quantum pigeonhole paradox without pre- and post-selection. Following this idea, we find a Hardy-like version of such paradox(referred to as the ``Hardy-like quantum pigeonhole(HLQP) paradox" in what follows), which can also be considered as a special kind of Hardy's paradox\cite{Hardy92,Hardy93}.

On the other hand, although the earliest version of Hardy's paradox\cite{Hardy92,Hardy93} can be considered as the simplest form of Bell theorem\cite{Mermin94},  a major shortcoming is that the success probability of excluding local realistic descriptions of quantum mechanics is not very high(the best record of previous versions is $1/2^{n-1}$\cite{Shuhanjiang2018}, where the qubit number $n\geq3$). This bottleneck may cast a gloom over the application of Hardy's paradox. How to improve that may depend on the investigation of some unconventional Hardy's paradoxes.  Since the HLQP paradox is a novel kind of Hardy's paradox, it may bring some unexpected advantages to address this problem.

In addition, recall that many quantum features can be exhibited vividly by their mathematical counterparts(for example, a kind of multi-particle entanglement can be represented by graphs\cite{Briegel2001,Briegel2004}). As far as we know, such elegant counterparts of a Hardy's paradox are very rare(another disadvantage). In order to fill this gap, an exploration for some mathematical counterparts to represent the HLQP paradox is necessary.

In this Letter, we  present a general construction of the $n$-qubit HLQP paradox. Besides, in order to give a suitable mathematical counterpart as a representation of the HLQP paradox, we introduce a kind of novel quantum state called the projected-coloring graph state, which can be represented by a kind of nontraditional graph called the projected-coloring graph. We show that each un-colorable projected-coloring graph can induce a HLQP paradox. Furthermore, we find that some $n$-qubit HLQP paradoxes(which are also $n$-qubit Hardy's paradoxes) can provide a greater success probability in showing Bell nonlocality(or contexuality) than previous Hardy's paradoxes($n>3$). In the end, we briefly discuss two other quantum paradoxes which are essentially equivalent to some HLQP paradoxes.

\section{A minimal Hardy-like quantum pigeonhole paradox}
 First, denote by $X_i~(Y_i,~Z_i)$ the Pauli matrices $\sigma_x~(\sigma_y,~\sigma_z)$ of the $i$-th qubit and  $|0\rangle$ and $|1\rangle$ the two eigenstates (with eigenvalues $+1$ and $-1$ respectively) of $Z$.
Besides, let two mutually orthogonal states, $|+\rangle=(|0\rangle+|1\rangle)/\sqrt{2}$ and $|-\rangle=(|0\rangle-|1\rangle)/\sqrt{2}$ be two ``boxes", and three qubits be three ``pigeons". Namely,  $X_i=1$ ($X_i=-1$) indicates that the $i$-th qubit(pigeon) is in the box $|+\rangle$ ($|-\rangle$). Here, $X_i=1$, for example, denotes the event that $X_i$ is measured and the outcome $+1$ is obtained.
To construct a HLQP paradox, one can usually assume that the quantum state admits a local hidden variable(LHV) model\cite{Bell,CHSH} or a noncontextual hidden variable model\cite{Bell2,KS,Mermin1}. For simplicity, hereafter we only discuss the HLQP paradox of ruling out the LHV model, and accordingly, we only consider the space-like separated measurements.

In analogy with the conventional quantum pigeonhole paradox, the minimal system of producing a HLQP paradox also requires three qubits. To begin with, we consider the three-qubit quantum state
\begin{align}\label{3qubit-hardy-state}
|\Psi\rangle=\alpha|\Phi\rangle+\beta|111\rangle,
\end{align}
where $|\alpha|^2+|\beta|^2=1~(|\alpha|\neq0)$ and
\begin{align}\label{3qubit-GHZ-LU}
|\Phi\rangle=\frac{1}{2}(|000\rangle-|011\rangle-|101\rangle-|110\rangle)
\end{align}
is a Greenberger-Horne-Zeilinger(GHZ) state\cite{GHZ1989,GHZ1990}, because
$|\Phi\rangle=(|\circlearrowleft\circlearrowleft\circlearrowleft\rangle+|\circlearrowright\circlearrowright\circlearrowright\rangle)/{\sqrt{2}}$. Here $|\circlearrowleft\rangle=(|0\rangle+i|1\rangle)/{\sqrt{2}}$ and $|\circlearrowright\rangle=(|0\rangle-i|1\rangle)/{\sqrt{2}}$.

Once the qubit $i$ is measured and found in the box $|0\rangle$ in a run of the experiment,  the other two qubits (``pigeons"), say $j$ and $k$, can be found in the state $|\phi^-_{ij}\rangle=\frac{1}{\sqrt{2}}(|0\rangle_i|0\rangle_j-|1\rangle_i|1\rangle_j)$. Namely, if $X_j$ and $X_k$ were assumed to be measured in this run, their values should satisfy $X_jX_k=-1$, which indicates that the qubits $j$ and $k$ cannot stay in the same box. Then one can get the following properties(referred to as ``Hardy-like constraints" thereinafter):
\begin{subequations}\label{3qubit-hardy-ph}
\begin{align}
  P(X_2X_3=-1|Z_1=1) &= 1,\label{3qubit-hardy-ph1} \\
  P(X_1X_3=-1|Z_2=1) &= 1,\label{3qubit-hardy-ph2}\\
  P(X_1X_2=-1|Z_3=1) &= 1, \label{3qubit-hardy-ph3}\\
  P(Z_1=1,Z_2=1,Z_3=1) &= \frac{|\alpha|^2}{4}. \label{3qubit-hardy-ph4}
\end{align}
\end{subequations}
Here $P(X_2X_3=-1|Z_1=1)$, for example,
is the conditional probability of $X_2$ and $X_3$ measured with outcomes satisfying $X_2X_3=-1$ given that the result of $Z_1=1$,
and $P(Z_1=1,Z_2=1,Z_3=1)$ stands for the joint probability of obtaining the result of $Z_1=1,Z_2=1,Z_3=1$.
Based on the constraints of Eqs.(\ref{3qubit-hardy-ph1}-\ref{3qubit-hardy-ph3}) and Eq.(\ref{3qubit-hardy-ph4}), one can construct a HLQP paradoxical argument as follows.

Consider a run of the experiment for which $Z_1,Z_2$ and $Z_3$ are measured and the results $Z_1=1,Z_2=1$ and $Z_3=1$ are obtained(it will happen with a probability of ${|\alpha|^2}/{4}$). Assume that the state admits a LHV model\cite{Bell,CHSH}. Since we have got $Z_1=1$, it follows from Eq.(\ref{3qubit-hardy-ph1}) that if $X_2$ and $X_3$ had been measured,  their results should satisfy $X_2X_3=-1$.  In fact, from the assumption of locality\cite{EPR}, even if $X_1$ had been measured on qubit 1, one would still have $X_2X_3=-1$. Thus in this run, the outcomes of measuring $X_2$ and $X_3$(determined by the hidden
variables $\lambda$) must satisfy $X_2(\lambda)X_3(\lambda)=-1$, indicating that qubits 1 and 2 cannot stay in the same box. Likewise, one can infer that $X_1(\lambda)X_3(\lambda)=-1$ and $X_1(\lambda)X_2(\lambda)=-1$ from the measurement results $Z_2=1$ and $Z_3=1$, respectively. Therefore, any two qubits cannot stay in the same box(for this run), which contradicts with classical pigeonhole principle(see also Appendix \ref{app-a}). Namely, a three-qubit HLQP paradox is produced. As a consequence, one can conclude that quantum correlations cannot be classically simulated and any realistic interpretations of quantum mechanics must be nonlocal.

{\it Remarks 1.---} (I) Some quantum state can induce more than one HLQP paradoxes(e.g. the results of $Z_1=Z_2=-1,Z_3=1$ in the above example can also induce a HLQP paradox, but it is equivalent to the case that replace $|\Phi\rangle$ with $\frac{1}{2}(|000\rangle+|011\rangle+|101\rangle-|110\rangle)$ and consider the results of $Z_1=Z_2=Z_3=1$. Similar trick also applies to the systems with more qubits). Without loss of generality, hereafter we mainly discuss the above type of HLQP paradox.
(II) Essentially, the HLQP paradoxical argument arises from the component $|\Phi\rangle$.  By contrast,
the component $|111\rangle$ is only used to ensure a greater universality rather than necessary. Besides, the above paradox with $|\alpha|=1$ can also be regarded as a Hardy's paradox induced from a three-qubit GHZ paradox\cite{GHZ1989,GHZ1990}.

\section{ A general $n$-qubit($n\geq3$) HLQP paradox}
 Inspired by the three-qubit HLQP paradox,  a general $n$-qubit HLQP paradox to be constructed is based on the following $n$-qubit state
\begin{align}\label{n-qubit-hardy-PH-state}
    |P\rangle=\alpha|A\rangle+\beta|B\rangle,
\end{align}
where $|\alpha|^2+|\beta|^2=1~(|\alpha|\neq0)$ and $\langle A|B\rangle=0$.
The normalized components $|A\rangle$ and $|B\rangle$ are defined as
\begin{subequations}\label{A-and-B}
\begin{align}
|A\rangle&=\frac{1}{\sqrt{p+1}}(|00\cdots0\rangle-\sum_{i\in\mI}\theta_{i}|\vec{0}\rangle_{\bar{\mS}_i}|\vec{1}\rangle_{\mS_i}),
\label{n-qubit-hardy-PH-key-state}\\
|B\rangle&=\sum_{i\in\mJ}\lambda_{i}|\vec{0}\rangle_{\bar{\mT}_i}|\vec{1}\rangle_{\mT_i},\label{hardy-PH-secondary-state}
\end{align}
\end{subequations}
where $\mS_i,\mT_i\subset\{1,2,\cdots,n\}$, $1\leq |\mS_i|<n$, $|\mS_{i}\cup\mS_{j}|>\max\{|\mS_{i}|,|\mS_{j}|\}$ ($i\neq j$) and $|\bar{\mS}_{r}\cap\bar{\mT}_{s}|<|\bar{\mS}_{r}|$ ($\forall r,s$). Besides, $|\vec{1}\rangle_{\mS_i}\equiv\otimes_{k\in\mS_i}|1\rangle_k$, $|\vec{0}\rangle_{\bar{\mS}_i}\equiv\otimes_{k\in\bar{\mS}_i}|0\rangle_k$. Moreover,  the index set $\mI$ is used to describe a group of specific subsets of $\{1,2,\cdots,n\}$, and $\mJ$ is used to indicate all possible $\mT_i$. The coefficients $\theta_{i}=\pm1$ and
$\lambda_{i}\in\mC$. Note that $p=\sum_{i\in\mI}|\theta_{\mS_i}|$.
In fact, the component $|A\rangle$ ($|B\rangle$) can be considered as a generalization of $|\Phi\rangle$ ($|111\rangle$) in Eq.(\ref{3qubit-hardy-state}).

For a given $|P\rangle$ (or $|A\rangle$), one can define a {\it Hardy matrix} $A$ and its {\it argumented Hardy matrix} $B$  as follows:
\begin{align}\label{Hardy-matrix-component}
  A=\left(
      \begin{array}{c}
        A_{ij} \\
      \end{array}
    \right)_{p\times n}, ~~~B=\left(
    \begin{array}{c|c}
      A& \vec{\Theta} \\
    \end{array}
  \right),
\end{align}
where the components $A_{ij}=1$ if $i\in\mI$ and $j\in\mS_i$; $A_{ij}=0$ otherwise.
Besides, the $i$-th component of $\vec{\Theta}$ is $\Theta_i=(\theta_{\mS_{i}}+|\theta_{\mS_{i}}|)/2$.

In a run of the experiment, if $Z_{j_1},Z_{j_2},\cdots,Z_{j_{n-|\mS_i|}}$ $(\{j_1,j_2,\cdots,j_{n-|\mS_i|}\}=\bar{\mS}_{i})$ are measured and the results $Z_{j_1}=Z_{j_2}=\cdots=Z_{j_{n-|\mS_i|}}=1$ are obtained, i.e., the qubits in $\bar{\mS}_i$ are found in $|\vec{0}\rangle_{\bar{\mS}_i}$,
then the qubits in $\mS_i$ are in the eigenstate of $\prod_{k\in\mS_{i}}X_{k}$
(with the eigenvalue $-\theta_i$). Therefore, for any $i\in\mI$,
\begin{align}\label{n-qubit-Hardy-QH1}
  P(\prod_{k\in\mS_{i}}X_{k}=-\theta_i|&Z_{j_1}=Z_{j_2}=\cdots=Z_{j_{n-|\mS_i|}}=1)=1,
\end{align}
Besides, Denote by $\cup_{i=1}^p\bar{\mS}_{i}=\{p_1,p_2,\cdots,p_q\}$. since $|\cup_{i=1}^p\bar{\mS}_{i}|\neq|\bar{\mT}_j|$ ($\forall j\in\mJ$) and $|\cup_{i=1}^p\bar{\mS}_{i}|>|\bar{\mS}_k|$ ($\forall k\in\mI$),  we can get
\begin{align}\label{n-qubit-Hardy-QH2}
  P(Z_{p_1}=Z_{p_2}=\cdots=Z_{p_q}=1)=\frac{|\alpha|^2}{p+1} >0.
\end{align}

Note that Eq.(\ref{n-qubit-Hardy-QH1}) and  Eq.(\ref{n-qubit-Hardy-QH2}) can provide a total of $p+1$ constraints. Then a practical criterion for constructing the HLQP paradox can be described as follows.

{\it Theorem 1.} --- Given an $n$-qubit state $|P\rangle=\alpha|A\rangle+\beta|B\rangle$ in which $|A\rangle$ and $|B\rangle$ are defined in Eqs.(\ref{n-qubit-hardy-PH-key-state}) and (\ref{hardy-PH-secondary-state}),
one can always construct a HLQP paradox  if $rank(A)\neq rank(B)$, where $A$ and $B$ are the corresponding Hardy matrix and argumented Hardy matrix, respectively.

{\it Proof.}--- See Appendix \ref{app-b}.\hfill$\blacksquare$

For the extreme case of $|\alpha|=1$, it is worthwhile to notice the following corollary.

{\it Corollary 1.} --- The $n$-qubit state $|A\rangle$  can always induce a HLQP paradox  if $rank(A)\neq rank(B)$.

Theorem 1 (or Corollary 1) can induce a formalized approach to construct a  HLQP paradox(which could facilitate the computer search). In some special cases,  one can even find several analytic constructions for the HLQP paradox.  For example,  a $(2k+1)$-qubit($k\geq1$) PCG state with $\mS_i=\{i,i+1\}$($i=1,2,\cdots,2k+1$, and $i+1=1$ if $i=2k+1$) and $\theta_i=-1$, can always produce a HLQP paradox, since $rank(A)=n-1$ and $rank(B)=n$.

Furthermore, the qudit version of HLQP paradox can also be constructed(see an example in Appendix \ref{app-c}).

\section{ A graphical representation of the HLQP paradox}
 Many well-known quantum features or phenomena may have some vivid descriptions in terms of their mathematical counterparts, but such elegant descriptions are rare for the Hardy's paradox. To fill the gap,  here we shall present a graphical representation of the HLQP paradox.

To start, let us take a look at some notions.

An $n$-qubit {\it projected-coloring graph(PCG) state} $|G_C\rangle$ is defined as a quantum state which takes the form of Eq.(\ref{n-qubit-hardy-PH-key-state}), i.e., $|G_C\rangle=|A\rangle$.

\begin{center}
\begin{figure*}[t]
  \centering
\includegraphics[scale=0.65]{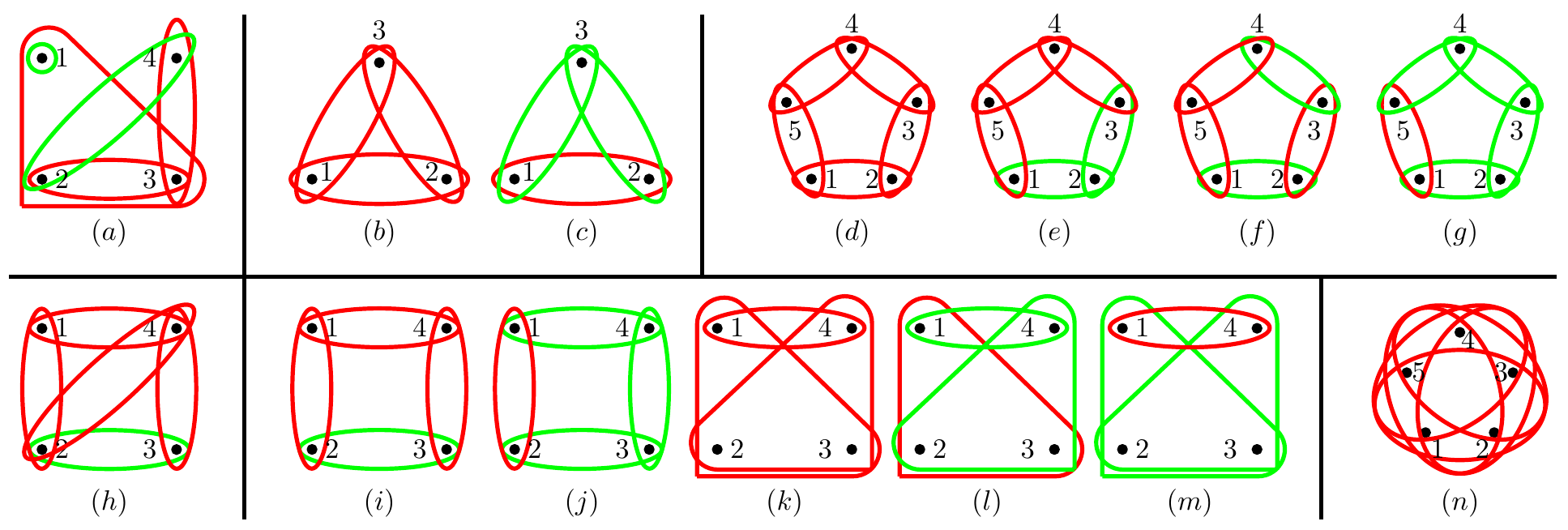} \caption{\label{hardy-ph-graph-eg} (a) An illegal PCG( $|\{1\}\cup\{1,2,3\}|=3$, $|\{2,3\}\cup\{1,2,3\}|=3$). (b)-(g) Some typical three-qubit and five-qubit irreducible un-colorable loop PCGs.  (h) A four-qubit PCG containing a three-qubit un-colorable sub-PCG. (i)-(m) Typical  four-qubit irreducible  un-colorable  PCGs. (n) Another five-qubit  un-colorable PCG.}
\end{figure*}
\end{center}

A {\it projected-coloring graph}  $G_C=(V,E)$ associated with the PCG state $|G_C\rangle$ can be defined as an unconventional weighted graph which consists of a set of vertices $V=\{1,2,\cdots,n\}$ and a set of edges $E=\{\mS_i|i=1,2,\cdots,p\}$ with  weights  red $(R)$ and green $(G)$ (corresponding to $\theta_{\mS_i}=1$ and  $\theta_{\mS_i}=-1$ respectively), where any two edges $\mS_i,\mS_j\in E$ should satisfy $|\mS_i\cup \mS_j|>\max\{|\mS_i|,|\mS_j|\}$ (each edge cannot be a subset of any other edges, see a counterexample in  Fig.\ref{hardy-ph-graph-eg}(a)). Besides, hereafter only connected PCGs(with no isolated sub-structures) are considered.

Next,  we consider such a vertex-coloring game: For a given PCG $G_c$, check whether there exists a consistent coloring scheme for all the vertices, wherein the coloring rules are described as follows.
\begin{itemize}
  \item (1) Each vertex $v_i$ can only be colored with either $R$ or $G$. If $v_i$ is colored by $R$, its coloring value $C(v_i)$ is defined as $C(v_i)=-1$; otherwise $C(v_i)=1$.
  \item (2) If the weight of the edge $\mS_i$ is $R$, its weight value $C(\mS_i)$ is defined as $C(\mS_i)=-1$;  otherwise $C(\mS_i)=1$. Namely $C(\mS_i)=-\theta_{\mS_i}$.
  \item (3) If there exist at least one vertex-coloring solution, such that $\prod_{v_i\in\mS_i}C(v_i)=C(\mS_i)$ holds for any $\mS_i\in E$, then the PCG $G_C$ is colorable; otherwise $G_C$ is un-colorable.
\end{itemize}

This game can be naturally associated to the HLQP paradox by the following theorem and corollary.

{\it Theorem 2.}--- There exists a one-to-one correspondence between an
un-colorable (colorable) PCG and the condition $rank(A)\neq rank(B)$ ($rank(A)=rank(B)$).

{\it Proof.} --- See Appendix \ref{app-d} for details. \hfill$\blacksquare$

{\it Corollary 2.}--- If $|A\rangle$ is associated with an un-colorable PCG, then $|P\rangle$ can induce a HLQP  paradox.

{\it Proof.}---  By combining  Theorem 1 and Theorem 2, one can complete the proof.  \hfill$\blacksquare$

Conversely,  any HLQP paradox can also be represented by a un-colorable PCG(not a one-to-one correspondence since $|P\rangle$ with the same $|A\rangle$ is not unique).

Moreover,  an $n$-qubit HLQP paradox is said to be a {\it genuinely $n$-qubit HLQP paradox} if one cannot reduce the number of Hardy-like constraints and still have a HLQP paradox, and the corresponding PCG is called an {\it irreducible un-colorable PCG}(several examples and a counterexample are shown in Fig.\ref{hardy-ph-graph-eg}).

{\it Remarks 2.---} (I) Un-colorable PCGs are also useful for the conventional quantum pigeonhole paradox(see Appendix \ref{app-e}). (II) Unlike the graph(or hypergraph) state\cite{Briegel2001,Briegel2004,Gachechiladze2016}, the PCG state is a ``conditional subsystem stabilizer state" rather than a stabilizer state, and sometimes this may bring us some unexpected advantages.

{\it Success probability}.--- For an $n$-qubit Hardy's Paradox,  the maximal success probability of excluding local realism in previous versions\cite{Cereceda04,Shuhanjiang2018} is $1/2^{n-1}$. Actually, this probability can be greatly improved by some HLQP paradoxes. For example, consider the HLQP paradox induced from the PCG state(associated with an $n$-vertex loop PCG) $|A_{L_n}\rangle=\frac{1}{\sqrt{n+1}}(|\vec{0}\rangle_V-|1\rangle_1|1\rangle_{n}|\vec{0}\rangle_{V\backslash \{1,n\}}+\sum_{i=1}^{n-1}|1\rangle_i|1\rangle_{i+1}|\vec{0}\rangle_{V\backslash \{i,i+1\}})$. Clearly, the success probability is $P_n^L=1/(n+1)$, which decays much slower over $n$ than $P_n^{G}$ and $P_n^{S}$ listed in Table \ref{TB1}. Namely, this HLQP paradox is more efficent\cite{successpro} in demonstrating Bell nonlocality than two representative Hardy's Paradoxes\cite{Cereceda04,Shuhanjiang2018}.

\begin{table}
\caption{The success probabilities of three kinds of Hardy's paradoxes for $n$ qubits. } \label{TB1}
 \begin{threeparttable}
\begin{tabular}{l|l}
  \hline\hline
\text{Scenarios} &  The success probabilities ($n\geq3$)\\
  \hline
{\text{Loop PCG state induced}} & $P_n^{L}={1}/{(n+1)}$ \\
{\text{Generalized, Ref.\cite{Shuhanjiang2018}}} & $P_n^{G}={1}/{2^{n-1}}$ \\
{\text{Standard, Ref.\cite{Cereceda04}}} & $P_n^{S}={1}/{2^n}\times(1+\cos\frac{\pi}{n-1})$ \\
\hline\hline
\end{tabular}
\begin{tablenotes}
\item[1] {For $n\geq3$, $P_n^{L}\geq P_n^{G} >P_n^{S}$.}
\end{tablenotes}
\end{threeparttable}
\end{table}

\begin{figure}
\includegraphics[width=3 in]{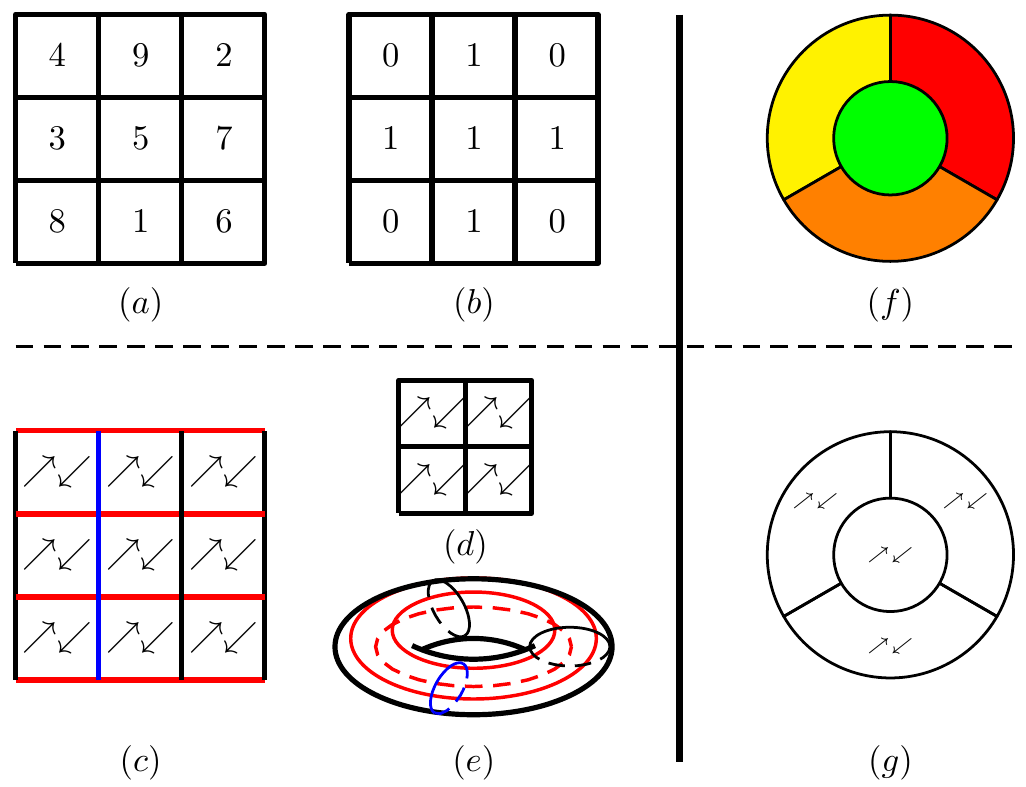} \caption{\label{magicsquare and four colors} (a) A traditional magic square of order 3, wherein the sum of the numbers is the same in each row, column, and
both diagonals. (b) A binary  magic square of order 3, in which the XOR sum of the numbers in each row, column, and both diagonals is the same. (c) A binary  magic squares associated with a $9$-qubit PCG state. (d) A $2\times2$ binary  magic square associated with a $4$-qubit PCG state. (e) A $3\times3$ binary magic square on a torus.
(f) A minimal map coloring example shows that at least four colors are required in the classical map coloring problem. (g) Specific measurements performed on a PCG state associated with this map can produce a Hardy-like quantum map coloring paradox.}
\end{figure}

\section{ Hardy-like quantum magic square paradox and Hardy-like map coloring paradox}
Besides the un-colorable PCGs, some other mathematical objects to mimic certain specific HLQP paradoxes can also be found, such as impossible magic squares and impossible maps, and the paradoxes induced from them are called ``Hardy-like quantum magic square paradox" and ``Hardy-like quantum map coloring paradox", respectively. Here we shall give two examples.

A {\it binary magic  square}(e.g. Fig.\ref{magicsquare and four colors}-(b)) is a generalization of a conventional magic square(e.g. Fig.\ref{magicsquare and four colors}-(a)), wherein the XOR sum of the numbers(zeroes or ones)  in each row, column, and both diagonals is the same, where XOR stands for addition modulo 2(denoted as $\oplus$). Associate the four-qubit PCG state $|M(4)\rangle=\frac{1}{\sqrt{7}}(|0000\rangle-|1100\rangle-|1010\rangle-|1001\rangle-|0110\rangle-|0101\rangle-|0011\rangle)$ to a $2\times2$ binary magic square(each grid stands for a qubit), see Fig.\ref{magicsquare and four colors}-(d).  Assume that $|M(4)\rangle$ can be modeled by LHV, then a HLQP paradox can be produced: Consider a run of the experiment for which $Z_1,Z_2,Z_3,Z_4$  are measured and the results $Z_1=Z_2=Z_3=Z_4=1$ are obtained. Similar to the discussion of the aforementioned HLQP paradoxes, one can get $X_1X_2=X_1X_3=X_1X_4=X_2X_3=X_2X_4=X_3X_4=-1$, a contradiction. On the other hand, let $m_k=(X_k+1)/2$ be the number arranged in the $k$-th grid of the binary magic square. Notice that
$X_iX_j=(-1)^{m_i\oplus m_j}=-1$($i\neq j$). It follows that $m_1\oplus m_2=m_1\oplus m_3=m_1\oplus m_4=m_2\oplus m_3=m_2\oplus m_4=m_3\oplus m_4=1$, giving rise to an impossible binary magic square. Then a Hardy-like quantum magic square paradox is produced. But for other binary magic squares, such as Fig.\ref{magicsquare and four colors}-(c) and (d),  extra constraints are required to induce such paradoxes(See Appendix \ref{app-f}).

on the other hand, to construct a Hardy-like quantum map coloring paradox, one can associate a quantum state to a map(each region stands for a qubit). For a run of the experiment, we use the value of $X_i$ of the $i$-th qubit(region) to label the ``color" of the $i$-th region. For example,  we associate $|M(4)\rangle$ to the map of Fig.\ref{magicsquare and four colors}-(g). Assume that $|M(4)\rangle$ admits a LHV  model. Likewise, if $Z_1,Z_2,Z_3,Z_4$ are measured in a run of the experiment and the outcomes $Z_1=Z_2=Z_3=Z_4=1$ are obtained,  one can infer that $X_1X_2=X_1X_3=X_1X_4=X_2X_3=X_2X_4=X_3X_4=-1$ according to local realism. Namely, in this run, two colors are enough to ensure each pair of the adjacent regions colored with different colors, which contradicts with the prediction of Four Color Theorem(Fig.\ref{magicsquare and four colors}-(f)). Then one can get a Hardy-like quantum map coloring paradox. Similar constructions also apply to more complicated maps.

\section{Discussion and conclusion}
To summarize, we have studied the general construction of the $n$-qubit HLQP paradox(which is also a special class of Hardy's paradox). Besides, by introducing the notions of the PCG state and the PCG, we give a pictorial representation of the HLQP paradox. This paradox has several advantages. From the angle of quantum pigeonhole paradox, the HLQP paradox scheme seems to be more friendly to the experimental physicists, and from the perspective of Hardy's paradox, a family of HLQP paradox has set a new record($1/(n+1)$) of the success probability of demonstrating Bell nonlocality(the record of the previous Hardy's paradoxes is $1/2^{n-1}$, see Ref.\cite{Shuhanjiang2018}).

Besides, some extended topics require further investigation, such as the qudit PCG state, the qudit HLQP paradox, and the analytic construction of some HLQP paradoxes. On the other hand, notice that any stabilizer of a system can exhibit a kind of symmetry. But usually a PCG state is only a ``conditional subsystem stabilizer state" rather than a full-system stabilizer state. The broken of such stabilizer symmetry may leads to some unknown properties and unexpected advantages. For example, ``all-versus-nothing" proofs\cite{all-vs-nothing} of Bell nonlocality which works for $100\%$ of the runs of an experiment(e.g. the GHZ paradox) are commonly induced from quantum states with perfect correlations(stabilizer states), but so far, such proofs induced from the systems without perfect correlations are still very rare.  By noticing that a PCG states may induce more than one HLQP paradoxes(see Remarks 1-(I)) and combining them together might give a more stronger demonstration of Bell nonlocality\cite{Cabello2001}, one could construct such all-versus-nothing proofs from some PCG states without perfect correlations\cite{WT2021}, which may have some potential applications in the field of information protection.

In addition, since every property of quantum mechanics not present in classical physics could lead to an operational advantage\cite{Resource2016,Resource2016A,Resource2017}, we believe that the HLQP paradox can also provide us some useful resources in certain quantum information processing.

\acknowledgments
We thank S. Ru, S. Paraoanu, F. Li  and N. Liu for useful suggestions and discussions.

\appendix
\renewcommand{\appendixname}{Appendix}
\section{Another way to get the contradiction of three-qubit HLQP paradox}\label{app-a}
If $|\Psi\rangle$ admits a LHV model, then Eqs.(\ref{3qubit-hardy-ph1}-\ref{3qubit-hardy-ph3}) in the main text imply that $\{Z_1=1\}\subset\{X_2X_3=-1\}$, $\{Z_2=1\}\subset\{X_1X_3=-1\}$ and $\{Z_3=1\}\subset\{X_1X_2=-1\}$. One can get $P(Z_1=1,Z_2=1,Z_3=1)\leq P(X_2X_3=-1,X_1X_3=-1,X_1X_2=-1)=0$, which contradicts with Eq.(\ref{3qubit-hardy-ph4}).

\section{Proof of Theorem 1}\label{app-b}

{\it Proof.}--- The proof is similar to the discussion of the three-qubit HLQP paradox.

Consider a run of the experiment for which $Z_{p_1},Z_{p_2},\cdots,Z_{p_q}$ are measured and the results $Z_{p_1}=1,Z_{p_2}=1,\cdots,Z_{p_q}=1$ are obtained, where $\{p_1,p_2,\cdots,p_q\}=\cup_{i=1}^p\bar{\mS}_i$.
For any $\mS_i$ ($1\leq i\leq p$), if $Z_{j_1},Z_{j_2},\cdots,Z_{j_n-|\mS_i|},X_{j_n-|\mS_i|+1},X_{j_n-|\mS_i|+2},\cdots, X_{j_n}$ are measured(where $\{j_1,j_2,\cdots,j_{n-|\mS_i|}\}=\bar{\mS}_i$, $\{j_n-|\mS_i|+1,j_n-|\mS_i|+2,\cdots,j_{n}\}={\mS}_i$), then necessarily $X_{j_n-|\mS_i|+1}X_{j_n-|\mS_i|+2}\cdots X_{j_n}=-\theta_{\mS_i}$ if $Z_{j_1}=Z_{j_2}=\cdots=Z_{j_n-|\mS_i|}=1$ by Eq.(\ref{n-qubit-Hardy-QH1}) in the main text.

Analogous to the argument of three-qubit HLQP prardox,
as long as the following conditions
\begin{align}\label{Hardy-QH-eqs1}
\left\{
  \begin{array}{ll}
    \prod_{k\in\mS_1}X_k=-\theta_{\mS_{1}}, & \\
    \prod_{k\in\mS_2}X_k=-\theta_{\mS_{2}}, &  \\
    \vdots, &  \\
   \prod_{k\in\mS_p}X_k=-\theta_{\mS_{p}}, &
  \end{array}
\right.
\end{align}
cannot hold simultaneously(namely, contradict with pigeonhole principle), a HLQP paradox can be produced. By performing logarithm operations on both sides of each equation over complex field, we can get
\begin{align*}
    \left\{
      \begin{array}{ll}
       \sum_{k\in\mS_1}\log X_k=\log (-\theta_{\mS_1}),& \\
       \sum_{k\in\mS_2}\log X_k=\log (-\theta_{\mS_2}),& \\
        \vdots, \\
       \sum_{k\in\mS_p}\log X_k=\log (-\theta_{\mS_p}).&
      \end{array}
    \right.
\end{align*}
Note that the value of each $X_k$ is either $+1$ or $-1$, $\log 1=0$ and $\log (-1)=\log(e^{i\pi})=i\pi$. Let $y_k=\ln X_k$ be the $k$-th component of vector $\vec{y}$, the above system of equations can be rewritten as
\begin{align*}
    A\vec{y}=i\pi\vec{\Theta},
\end{align*}
where $A$ is the Hardy matrix defined in Eq.(\ref{Hardy-matrix-component}) of the main text.
This system has at least one solution only if $rank(A)=rank\left(
                                                                    \begin{array}{c|c}
                                                                      A & i\pi\vec{\Theta} \\
                                                                    \end{array}
                                                                  \right)=rank\left(
                                                                    \begin{array}{c|c}
                                                                      A & \vec{\Theta} \\
                                                                    \end{array}
                                                                  \right)=rank(B)$.
Therefore, if  $rank(A)\neq rank(B)$, the system of equations (\ref{Hardy-QH-eqs1}) has no solution. In this case, a HLQP paradox  can be constructed. \hfill $\blacksquare$

{Remarks.}--- Generally speaking,  the above argument works only for a particular choice of the boxes($|+\rangle$ and $|-\rangle$).  If the boxes are not prescribed, then even $rank(A)=rank(B)$ holds, sometimes one can still construct a HLQP paradox.

A simple example is the three-qubit HLQP paradox based on $|\Psi\rangle$ with $\alpha=1$. If one choose the state $|\Phi^{\prime}\rangle=\frac{1}{2}(|000\rangle+|011\rangle+|101\rangle+|110\rangle)$ instead of $|\Phi\rangle$, one can get $rank(A)=rank(B)$. Let two boxes be $|\circlearrowleft\rangle=(|0\rangle+i|1\rangle)/\sqrt{2}$ and $|\circlearrowright\rangle=(|0\rangle-i|1\rangle)/\sqrt{2}$, a HLQP paradox  can still be constructed.  One can see that from the following constraints,
 \begin{subequations}\label{3qubit-hardy-ph-y}
\begin{align}
  P(Y_2Y_3=-1|Z_1=1) &= 1,\label{3qubit-hardy-ph-y1} \\
  P(Y_1Y_3=-1|Z_2=1) &= 1,\label{3qubit-hardy-ph-y2}\\
  P(Y_1Y_2=-1|Z_3=1) &= 1, \label{3qubit-hardy-ph-y3}\\
  P(Z_1=1,Z_2=1,Z_3=1) &= \frac{1}{4}>0.\label{3qubit-hardy-ph-y4}
\end{align}
\end{subequations}

In this scenario, one can also present a similar theorem based on such kind of boxes(by redefining the Hardy matrix $A$ and its augmented matrix $B$). In fact, this paradox is locally unitary equivalent to the three-qubit HLQP paradox introduced in the main text, and such equivalence also applies to more complicated cases. Therefore, it is enough to discuss the HLQP paradox in one choice of the boxes.

\section{An example of 4-qutrit HLQP Paradox}\label{app-c}

As a generalization of Pauli operators $\sigma_z$ and $\sigma_x$, two $d$-dimensional Pauli operators, $Z_d$ and $X_d$, can be defined as follows(without confusion, they are also written by $Z$ and $X$ for simplicity).
\begin{align*}
    Z=\sum_{n=0}^{d-1}e^{in\theta}|n\rangle\langle n|=\left(
                                             \begin{array}{ccccc}
                                               1 &  &  &  &  \\
                                                & e^{i\theta} &  &  &  \\
                                                &  & e^{2i\theta} &  &  \\
                                                &  &  & \ddots &  \\
                                                &  &  &  & e^{(d-1)i\theta} \\
                                             \end{array}
                                           \right),
\end{align*}
where $\theta=\frac{2\pi}{d}$.
\begin{align*}
    X=\sum_{n=0}^{d-1}|n\rangle\langle n\oplus1|=\left(
                                                    \begin{array}{cccccc}
                                                      0 & 1 &  &  &  &  \\
                                                       & 0 & 1 &  &  &  \\
                                                       &  & 0 & 1 &  &  \\
                                                       &  &  & \ddots & \ddots &  \\
                                                       &  &  &  & \ddots & 1 \\
                                                     1 &  &  &  &  & 0 \\
                                                    \end{array}
                                                  \right),
\end{align*}
where $a\oplus b\equiv (a+b)\mod d$, namely, $|d\rangle\equiv|0\rangle$. We have
\begin{align*}
    XZ=ZXe^{i\theta}.
\end{align*}

For $d=3$,  we can construct a 4-pigeon-3-pigeonhole  contradiction based on the following state
\begin{align*}
    |\Psi\rangle_{1234}=&{\frac{1}{3}}(|0000\rangle\cr
    &+\omega|0111\rangle+\omega|1011\rangle+\omega|1101\rangle+\omega|1110\rangle\cr
    &+\omega^2|0222\rangle+\omega^2|2022\rangle+\omega^2|2202\rangle+\omega^2|2220\rangle),\cr
\end{align*}
where $\omega=e^{i\frac{2\pi}{3}}$.
Then one have the following properties
 \begin{subequations}\label{4qutrit-hardy-ph-y}
\begin{align}
P(X_2X_3X_4=&\omega|Z_1=1)=1,\label{4qutrit-hardy-ph-y1}\\
P(X_1X_3X_4=&\omega|Z_2=1)=1,\label{4qutrit-hardy-ph-y2}\\
P(X_1X_2X_4=&\omega|Z_3=1)=1,\label{4qutrit-hardy-ph-y3}\\
P(X_1X_2X_3=&\omega|Z_4=1)=1,\label{4qutrit-hardy-ph-y4}\\
P(Z_1=Z_2=Z_3=&Z_4=1)=\frac{1}{9}.\label{4qutrit-hardy-ph-y5}
\end{align}
\end{subequations}
Consider a run of the experiment that measurements of $Z_1,Z_2,Z_3$,and $Z_4$ are performed(note that $Z_1,Z_2,Z_3$,and $Z_4$ are all unitary operators, namely, each $Z_j$ can also be written as $e^{iH_j}$, where $H_j$ is Hermitian. Then
measuring $Z_j$ can be converted to measuring $H_j$), and the results $Z_1=1,Z_2=1,Z_3=1$ and $Z_4=1$ are obtained. According to local realism,
if $X_1,X_2,X_3,X_4$ were measured in this run, then their results should satisfy $X_2X_3X_4=\omega,X_1X_3X_4=\omega,X_1X_2X_4=\omega$ and $X_1X_2X_3=\omega$, which contradicts with pigeonhole principle. Namely, we get a 4-qutrit HLQP Paradox.

Likewise,  this model can be generalized to the cases of $(d+1)$ qudits.

\section{Proof of Theorem 2}\label{app-d}

{\it Proof.} --- Notice that the coloring value of the $i$-th vertex  $C(v_i$) corresponds to the value assigned to $X_i$ of the $i$-th qubit.
Besides, the un-colorable condition states that $\prod_{v_i\in\mS_i}C(v_i)=-\theta_{\mS_i}$ cannot hold for all the edges, which corresponds exactly to $\prod_{k\in\mS_i}X_k=-\theta_{\mS_i}$. As mentioned in the proof of Theorem 1,  the condition of no solution for the system of equations $\prod_{k\in\mS_i}X_k=-\theta_{\mS_i}$ ($1\leq i\leq p$) is equivalent to $rank(A)\neq rank(B)$, we can then get one-to-one correspondence stated in Theorem 2. \hfill$\blacksquare$

\section{A supplementary conventional quantum pigeonhole paradox for three qubits}\label{app-e}

Sometimes un-colorable PCGs are helpful in looking for new conventional quantum pigeonhole paradoxes. For example, Fig.1-(c) in maintext
can induce another type of quantum pigeonhole paradox, while Fig.1-(b) corresponds to the original quantum pigeonhole paradox\cite{Aharonov2016}.

The initial state is prepared in $|\Phi_i\rangle=|+\rangle|-\rangle|+\rangle$, and  post-selected by $|\Phi_f\rangle=|0\rangle|1\rangle|0\rangle$, then one can check that $\langle\Phi_i|\frac{I+Y_1Y_2}{2}|\Phi_f\rangle=\langle\Phi_i|\frac{I-Y_1Y_3}{2}|\Phi_f\rangle=\langle\Phi_i|\frac{I-Y_2Y_3}{2}|\Phi_f\rangle=0$. It follows that at intermediate times the pair of qubits $\{1,3\}$ and $\{2,3\}$ should be put in the same box but $\{1,2\}$ should not, a contradiction according to the classical pigeonhole principle.

\section{More examples of Hardy-like quantum magic square
paradoxes}\label{app-f}
{\it Example 1}.--- Consider a $(4\times 4)$-qubit PCG state $|M(16)\rangle=\frac{1}{\sqrt{11}}(|\vec{0}\rangle_{\mS}-\sum_{i=1}^{10}|\vec{1}\rangle_{\mS_{i}}|\vec{0}\rangle_{\bar{\mS}_i})$, where $\mS=\{1,2,\cdots,16\}$ and $\{\mS_i|i=1,2,\cdots,10\}=\{\{4k+1,4k+2,4k+3,4k+4\}|k=0,1,2,3\}\cup\{\{l,4+l,8+l,12+l\}|l=1,2,3,4\}
\cup\{1,6,11,16\}\cup\{4,7,10,13\}$.
Assume that $|M(16)\rangle$ can be modeled by LHV. Consider a run of the experiment for which $Z_1,Z_2,\cdots,Z_{16}$  are measured and the results $Z_1=Z_2=\cdots=Z_{16}=1$ are obtained. Similar to the argument of the  HLQP paradox in the main text, one can finally conclude that $\prod_{j\in\mS_1}X_j=\prod_{j\in\mS_2}X_j=\cdots=\prod_{j\in\mS_{10}}X_j=-1$. Based on that,  one can find some solutions for $X_1,X_2,\cdots,X_{16}$. There is no contradiction.

Notice that $\prod_{i=1}^{10}(\prod_{j\in\mS_i}X_j)=X_1X_4X_6X_7X_{10}X_{11}X_{13}X_{16}=1$. One can consider another PCG state $|\tilde{M}(16)\rangle=\frac{1}{\sqrt{12}}(\sqrt{11}|M(16)\rangle-|1001011001101001\rangle)$. Namely, we impose a new conditional constraint:
If $Z_2=Z_3=Z_5=Z_8=Z_9=Z_{12}=Z_{14}=Z_{15}=1$ are obtained, then necessarily $X_1X_4X_6X_7X_{10}X_{11}X_{13}X_{16}=-1$. Next, we also consider a run of the experiment for which $Z_1,Z_2,\cdots,Z_{16}$  are measured and the results $Z_1=Z_2=\cdots=Z_{16}=1$ are obtained. Then this extra constraint ensures that there is no consistent solution for $X_1,X_2,\cdots,X_{16}$ in classical world according to pigeonhole principle.

Let $m_r=(X_r+1)/2$ be the number arranged in the $r$-th grid of the binary magic square.
Notice that $\prod_{j\in\mS_i}X_{j}=(-1)^{\oplus_{j\in\mS_i}m_j}=-1$ and
$X_1X_4X_6X_7X_{10}X_{11}X_{13}X_{16}=(-1)^{m_1\oplus m_4\oplus m_6\oplus m_7\oplus m_{10}\oplus m_{11}\oplus m_{13}\oplus m_{16}}=-1$.
It follows that $\oplus_{j\in\mS_1}m_j=\oplus_{j\in\mS_2}m_j=\cdots=\oplus_{j\in\mS_{10}}m_j=m_1\oplus m_4\oplus m_6\oplus m_7\oplus m_{10}\oplus m_{11}\oplus m_{13}\oplus m_{16}=1$, a contradiction(the assumption of local realism can ``induce" a binary magic square which is forbidden in classical world). Then we can get a $4$-order conditional(with an extra constraint) Hardy-like quantum magic square paradox.

{\it Remarks.}--- Commonly, there are some prescribed constraints for a classical magic square(e.g. the $3\times3$ conventional magic square is arranged with numbers $1,2,\cdots,9$). Even for a binary magic square, usually the number of zeros(or ones) to be arranged should be prescribed(e.g. $4$ zeroes and $5$ ones for a $3\times3$ binary magic square). However, for a quantum binary magic square, we would like to choose some other constraints, such as the extra constraint imposed in the above example. After all, our goal is just to show that a classically impossible magic square might be probabilistically produced if the associated quantum state admits a LHV model.

{\it Example 2}.--- Consider a nine-qubit PCG state $|M(9)\rangle=\frac{1}{3}(|000000000\rangle-|111000000\rangle-|0001110000\rangle-|000000111\rangle
-|100100100\rangle-|010010010\rangle-|001001001\rangle-|100010001\rangle-|001010100\rangle)$.
Assume that $|M(9)\rangle$ can be modeled by LHV. Consider a run of the experiment for which $Z_1,Z_2,Z_3,Z_4,Z_5,Z_6,Z_7,Z_8,Z_9$  are measured and the results $Z_1=Z_2=Z_3=Z_4=Z_5=Z_6=Z_7=Z_8=Z_9=1$ are obtained. Likewise, one can conclude that the relations $X_1X_2X_3=X_4X_5X_6=X_7X_8X_9=X_1X_4X_7=X_2X_5X_8=X_3X_6X_9=X_1X_5X_9=X_3X_5X_7=-1$ should be satisfied. There is also no contradiction.

Notice that the  product of these above eight relations gives rise to $X_1X_3X_7X_9=1$. One can use another PCG state $|\tilde{M}(9)\rangle=\frac{1}{\sqrt{10}}(3|M(9)\rangle-|101000101\rangle)$ to construct a HLQP paradox. Likewise,  consider a run of the experiment for which $Z_1,Z_2,Z_3,Z_4,Z_5,Z_6,Z_7,Z_8,Z_9$  are measured and the results $Z_1=Z_2=Z_3=Z_4=Z_5=Z_6=Z_7=Z_8=Z_9=1$ are obtained. Besides $X_1X_2X_3=X_4X_5X_6=X_7X_8X_9=X_1X_4X_7=X_2X_5X_8=X_3X_6X_9=X_1X_5X_9=X_3X_5X_7=-1$, one can get an extra relation $X_1X_3X_7X_9=-1$.
All such relations contradict with pigeonhole principle.

Let $m_k=(X_k+1)/2$ be the number arranged in the $k$-th grid of the binary magic square. One can get
$m_1\oplus m_2\oplus m_3=m_4\oplus m_5\oplus m_6=m_7\oplus m_8\oplus m_9=m_1\oplus m_4\oplus m_7=m_2\oplus m_5\oplus m_8=m_3\oplus m_6\oplus m_9=m_1\oplus m_5\oplus m_9=m_3\oplus m_5\oplus m_7=1$ and $m_1\oplus m_3\oplus m_7\oplus m_9=1$, which cannot hold simultaneously in classical world.  This contradiction can induce another conditional Hardy-like quantum magic square paradox.

Note: One can also consider the case that a magic square stays on a torus(e.g. Fig.2-(e) in the main text), and one may construct a similar Hardy-like quantum magic square paradox(under some extra constraints).

{\it Example 3}.--- We generalize the notion of binary magic square to the $n$-dimensional case. For example, a $3$-dimensional binary magic square of order 2 is  an arrangement of $k$ ones and $2^3-k$ zeros in a $2\times2\times2$-cube, such that the XOR sum of the numbers  in each edge, four main diagonals, and twelve other diagonals is the same.

Consider an eight-qubit PCG state $|M(8)\rangle=\frac{1}{\sqrt{C_8^2+1}}(|00000000\rangle-\sum_{i=1}^{C_8^2}|11\rangle_{\mS_i}|000000\rangle_{\bar{\mS}_i})$, where $\mS_i=\{a_i,b_i\}$ and $a_i\neq b_i\in\{1,2,3,\cdots,8\}$. Also assume that $|M(8)\rangle$ can be modeled by LHV. Consider a run of the experiment for which $Z_1,Z_2,\cdots,Z_{8}$  are measured and the results $Z_1=Z_2=\cdots=Z_{8}=1$ are obtained. Likewise, one can finally conclude that $\prod_{j\in\mS_1}X_j=\prod_{j\in\mS_2}X_j=\cdots=\prod_{j\in\mS_{28}}X_j=-1$, which contradict with pigeonhole principle.

Let $m_r=(X_r+1)/2$.  Notice that $\prod_{j\in\mS_i}X_{j}=(-1)^{\oplus_{j\in\mS_i}m_j}=-1$. Then one can get
$\oplus_{j\in\mS_1}m_j=\oplus_{j\in\mS_2}m_j=\cdots=\oplus_{j\in\mS_{28}}m_j=1$, a contradiction. Namely, we get a generalized Hardy-like quantum magic square paradox.


\begin{thebibliography}{30}
\bibitem [{suc()}]{successpro}%
  \BibitemOpen
  \href@noop {} {}\bibinfo {note} {Note that there also exist HLQP paradoxes
  which can seemingly provide a greater success probability than $P_n^L$. For
  example, the HLQP paradox induced from
  $\frac{1}{2}(|\vec{0}\rangle_V-|1\rangle_1|1\rangle_2|\vec{0}\rangle_{V\backslash\{1,2\}}-|1\rangle_1|\vec{0}\rangle_{V\backslash
  \{1\}}-|1\rangle_2|\vec{0}\rangle_{V\backslash \{2\}})$. The success
  probability is $P_n^T=1/4$(constant). But this class of the HLQP paradox can
  be proven to be equivalent to a three-qubit HLQP paradox.}\BibitemShut
  {Stop}%
\end{thebibliography}
\end{document}